\documentstyle[12pt]{article}
\begin{document}
\def \P{ {P_{T} }}
\def \d {\delta x}
\def  \dx {( x(t +T) - x(t) )}
\newcommand{\ov}{\overline}

\begin{center}
\centering{\bf Scaling and correlation in financial data}
\vskip 0.5cm
\end{center}

\begin{center}
\centering{  Rama CONT $^{1,2,3}$ 
\vskip 2cm

{\it 1) Service de Physique de l'Etat Condens{\'e}\\
 CEA Saclay\\
91191 Gif sur Yvette , France}
\vskip 0.5cm
{\it 2) Laboratoire de Physique de la Mati\`ere Condens{\'e}e\\
 CNRS URA 190 Universit\'e de Nice, 06108 Nice, France}
\vskip 0.5cm
{\it 3) Science \& Finance SA\\
 109-111 rue V. Hugo, 92 532 Levallois, France.}\footnote{Address for correspondence}
\vskip 0.5cm
{\it E-mail : cont@ens.fr }
\vskip 0.5cm
Fax : + 33 - 1 - 41 27 03 50
}
\end{center}
PACS classification numbers:\\
\noindent
02.50  Probability theory, stochastic processes and statistics\\
05.40.+j  Fluctuation phenomena, stochastic processes and statistics\\
89.90+n  Other areas of general interest to physicists
\vskip 1cm
Section: {\it Cross-Disciplinary Physics}.
\newpage
\vskip 1cm
\begin{abstract}
 
We apply the concept of scaling to the study of some statistical properties
of financial data, showing the link between scaling behavior on
one hand and  correlation properties on the other hand.
A non-parametric approach is used to study the scale dependence
of the empirical distribution of the price increments $\dx$ 
of S\&P Index futures, for  time scales  ranging from a few minutes
to a few days
using high-frequency price data. We show that
while the variance increases linearly with the timescale,
the kurtosis exhibits anomalous scaling properties, indicating a departure
from the iid hypothesis.
Study of the dependence structure of the increments
shows that although the autocorrelation function decays rapidly to zero
in a few minutes, the correlation of their squares exhibits a slow power law decay with exponent $\alpha \simeq$ 0.37, indicating persistence in the scale of fluctuations.
We establish a link between the scaling behavior and the dependence
structure  of the increments : in particular it is shown that the anomalous scaling
of kurtosis may be explained by "long memory" properties  of the square of the increments.

\vskip 1cm

Nous etudions les propri\'et\'es statistiques  des fluctuations 
des prix boursiers en utilisant le concept de lois d'\'echelle, 
en mettant en valeur le lien entre lois d'echelle et corr\'elations
temporelles
des incr\'ements de prix.
  Utilisant 
des donn\'ees haute-fr\'equence nous \'etudions, 
avec une approche non-param\'etrique, la mani\`ere
 dont   la distribution $\P$ des incr\'ements de prix varie avec l'\'echelle de temps 
$T$ \`a travers la d\'ependance de sa variance et sa kurtosis par rapport \`a   l'\'echelle de temps $T$, pour $T$ allant de quelques minutes \`a plusieurs jours.
Alors que la variance
cro\^{\i}t lin\'eairement avec l'\'echelle de temps, le kurtosis manifeste une
d\'ecroissance plus lente que celle d'une marche al\'eatoire \`a
incr\'ements ind\'ependants, indiquant la pr\'esence d'une d\'ependance non-lin\'eaire dans les incr\'ements, ce que confirme l'\'etude
de fonctions de corr\'elation non-lin\'eaires, r\'ev\'elant une autocorr\'elation
positive 
du carr\'e des fluctuations decroissant en loi de puissance avec un exposant
$\alpha \simeq$ 0.37.
Nous mettons \'egalement en \'evidence le lien entre 
cette persistance du carr\'e des incr\'ements d'une part et la decroissance lente de la kurtosis avec l'\'echelle de temps d'autre part.
  
\end{abstract}

\newpage

\section{Motivation} 
 
The distributional properties of asset price increments are important
both from a theoretical point of view, for understanding market dynamics,
 and for numerous applications, such as the pricing of derivative products 
and Value-at-Risk estimations,
in which distributional assumptions play a crucial role. These applications,
and many others, involve different time horizons, ranging from a few minutes,
the typical timescale on which market transactions take place, to several months
which is the time horizon portfolio managers are concerned with. This requires 
knowing the distributional properties of price increments at various time scales.

The problem of comparing the distributions of price changes at various time
scales is also interesting from a fundamental point of view. Indeed, 
as we explain below, whereas studying the distribution of increments
on a single timescale cannot distinguish between a process with independent increments
and one with dependent increments, 
studying the deformation of the distribution
under a change of time scale provides insight into the dependence
structure of the time series.
We will further elaborate on this point in section 5.
 
Many studies have been conducted on the distributional properties of asset prices
 and returns (for a review see \cite{pagan}).  However  most of these
studies focus on the distributional
properties of returns
\footnote{Most studies focus on the statistical properties of 
{\it returns}, defined as $\dx$ where $x(t)$ is the logarithm of the price. 
In this study however, we have chosen to focus instead on the properties of price increments : 
$x(t)$ is taken to be the price of the asset at date $t$.}
for a given value of the timescale $\tau$, using daily, weekly
or monthly data for example.

Mandelbrot was  the first to emphasize the idea of comparing the 
distributional properties of price changes 
at different time scales \cite{mandelbrot1}.
The idea behind Mandelbrot's approach is that of {\it scale invariance}: 
 the distribution ${P_T}$ of price changes
on a time scale $T$ may be obtained from that of a shorter time scale $\tau < T$  by an appropriate rescaling of the variable:

$$
P_T(x) = \frac{1}{\lambda} P_\tau(\frac{x}{\lambda}) \ , \  \ \lambda = (\frac{T}{\tau})^H
$$

where $H$ is the self-similarity exponent. The simplest scenario of a scale
invariant price process is that of a  {\it L\'evy flight}: a random walk with {\it iid} increments
having a stable L\'evy distribution\cite{mandelbrot8,gnedenko}. This was indeed the solution proposed
by Mandelbrot \cite{mandelbrot1}. In this case, if the increments 
have a $\mu$-stable L\'evy distribution then the process is self-similar 
with exponent $H= 1/\mu$.

In the recent years the availability of high frequency financial data has
rendered possible the study of price dynamics over a wider range of time
scales. Several studies  have been done in a spirit similar
to that of Mandelbrot \cite{mandelbrot1} in the quest of self-similarity
in financial time series,  on the S\&P 500 index
\cite{stanley} and on the CAC40 (Paris stock index) \cite{zaj}.
This scale invariant behavior, observed for short time scales (up to
1000 minutes in \cite{stanley}), breaks down for longer time scales \cite{houches}. Explanations for these observations have been given
in terms of {\it truncated L\'evy flight} models\cite{houches,tlf}.

While inspired by the above approaches, the present study has a different aim:
we argue that the existence of non-linear correlations and anomalous scaling
indicate that simple random walk models may be insufficient for modeling
finer aspects of price fluctuations. 
Without assuming any scale {\it invariance} or self-similarity property,
 we attempt to characterize  the deformation of the probability density function ${P_T}$ under a change of time scale $T$  
by studying the {\it scaling behavior} 
of its variance and its kurtosis, showing that their scaling properties
deviate from that of an {\it iid} random walk. 

Knowledge of the scale dependence of distributional properties
of price changes enables us to extract more information on the dynamics of 
series than  studying it on a single time scale, providing information
about the dependence structure of the increments : in particular, we show that
a link can be established between the scaling behavior of the cumulants and 
high-order correlation functions of the time series.

\section{The data set}

The study has been conducted on a high frequency data set of
S\&P Index future prices between October
1991 and September 1995. There are four maturity dates
each year for S\&P futures contracts and market activity is most intense
for the contract closest to its maturity. We have constituted our data set
by recording for each period the price of the S\&P futures contract
which has the greatest liquidity
among all contracts traded in the market, using the above criterion:
$x(t)$ is the price of the (unique) futures contract maturing less than three months
from $t$. The procedure of  sticking together various
price series may artificially introduce price jumps  at  the
maturity dates. The presence of these jumps may in turn influence
various statistical estimates such as the variance or kurtosis.  
To avoid this, we have shifted all prices in the next data set
by a constant so as to equalize the last price of a given set with the
first price of the next one, thus eliminating these jumps.  
Prices of adjacent contracts are shifted by a constant value such that there
 is no jump between adjacent  prices at  maturity dates.
Since our study focuses on statistical properties of price {\it differences},
shifting adjacent prices by the same constant has no effect whatsoever on our results.

Prices are recorded at 5 minute intervals,
the price retained being the last price recorded during 
the corresponding 5-minute interval. The data set thus constituted
contains more than 75 000 prices.  Throughout the article, time is given
in multiples of $\tau =$ 5 minutes ( $T = N\tau$). N = 84  corresponds to
a working day, N=420 to a regular working week.

The size of our data set enables us to study timescales ranging from 5 minutes
(N=1) up to a few days ( N= 200). Estimates for the longest time scale (N=200)
thus correspond  to averages over a sample of size $\geq$ 3500, enabling 
large sample approximations to be used. 

\section{Probability density function of the increments}

Let $K$ be a smooth, positive function with the property
$$
\int_{-\infty}^{+\infty} K(x) dx = 1 
$$

If $(X_i)_{n\geq i \geq 1}$ is  a sample realization of a random variable 
$X$ the kernel estimator \cite{silverman} defined by
$$
f(x) = \frac{1}{nh} \sum_{i=1}^{n} K(\frac{x-X_i}{h})
$$
gives a smooth estimator of the probability density function (PDF)
of the variable X. We have calculated a kernel estimator for the PDF
of the distribution of the 5 minute price changes using a Gaussian kernel i.e. $K(x) = \frac{\exp{(-x^2/2)}}{\sqrt{2\pi}}$, shown in Figure 1.
The  non-normal character of the distribution is better illustrated
when it is compared to a Gaussian curve with same mean and variance, also
shown in Figure 1.

This impression is confirmed upon calculating the excess kurtosis $\kappa_{5min}$ of the distribution (see section 4): $\kappa = 15.95$.
This leptokurtic character of price changes is a constant feature of high frequency data.

Another feature of high frequency data is that drift effects  are negligible
 compared to fluctuations i.e. the ratio of the mean to the standard deviation
is very small. This ratio  is found to be   $m/\sigma \simeq 3.3\ 10^{-3}$ in our case:
price variations due to drift are roughly a thousand times smaller than those due to 'volatility' - fluctuations around the mean value $m$. Therefore for  practical
purposes the distribution may be considered as centered around zero as long as the time scales $T$ considered are such that $mN \ll \sigma \sqrt{N}$ i.e. $ N\ll N_0 = (\sigma/m)^2  = 9000   \simeq $ 3 months.

\section{Scaling properties}

We calculate for each time scale $T=N\tau$ the following quantities:

\begin{eqnarray*}
\sigma(T)^2 & = &  \ov{[\dx  -  \ov{\dx} ]^2} \\
\kappa(T) & = & \frac{\ov{ [\dx - \ov{\dx}]^4 } }{\sigma(T)^4} - 3  
\end{eqnarray*}

where the symbol $\ov{( . )}$ denotes a sample average:

$$
\ov{ A } = \frac{1}{n}\sum_{t=1}^{n} A( t )
$$

$\sigma(T)$ is an estimator of the variance of the distribution  $\P$ and 
$\kappa(T)$ an estimator of the {\it kurtosis}, 
the kurtosis of a random variable $X$ being defined as:

$$
\kappa_X = \frac{E[ ( X - E(X) )^4 ] }{E[ ( X - E(X) )^2 ]^2 }- 3
$$

(where $E$ denotes expectation value)
defined such that $\kappa_X = 0$ for a gaussian random variable.

Going from a shorter time scale $\tau$ to a longer time scale $N\tau$
 formally corresponds  to  summing $N$ random variables:

$$
x(t+N\tau) - x(t) = \sum_{j=1}^{N} \d_j
$$

If  increments are stationary and independent, the distributions for various timescales
are simply related by a convolution relation: 

$$P_{N\tau} = P_{\tau} \otimes  P_{\tau} \otimes ... \otimes  P_{\tau}$$

where $\otimes$ denotes convolution.One can then deduce simply from this relation that
for a stochastic process with {\it iid} increments with finite variance,
 the variance of the distribution at scale $T = N\tau$ increases linearly with $N$
and that the kurtosis decreases as $1/N$ \cite{feller}. More generally if $c_k(N)$ is the k-th 
semi-invariant or cumulant \cite{gnedenko,feller} of the distribution $\P = P_{N\tau}$ then the normalized
cumulants

$$\lambda_{k}(N\tau) = \frac{c_k(N\tau)}{\sigma(N\tau)^{k}} = \frac{\lambda_{k}(\tau)}{N^{k/2  -  1}}$$

 tend to zero for large N, which is another way of seeing the Central Limit Theorem: the
cumulants of order $\geq 3$ of the limit distribution are all zero, a property which 
characterizes the Gaussian distribution. For example the skewness $\lambda_3(N\tau)$
decreases as $1/\sqrt{N}$. Note that $\lambda_4(N\tau)$ is the kurtosis $\kappa(N\tau)$
of the distribution $P_{N\tau}$.

\noindent
There is no  {\it a priori} reason  to believe  that the price of an asset has independent
increments. However, it is now known that the independence of increments
is not a necessary condition for the Central Limit Theorem to apply \cite{taqqu}:
some types of dependence structures still allow for convergence to the Gaussian
distribution as $T\rightarrow\infty$ in which case the distribution will resemble
more closely a Gaussian at longer time scales than at shorter ones and 
$ \lambda_{k}(T) \rightarrow 0$ as $T\rightarrow\infty$. However there is no  reason anymore for the normalized cumulants  $\lambda_k(N\tau)$ to decrease as $1/{N^{k/2  -  1}}$ as in the {\it iid} case: the presence of  
'non-trivial'  scale dependence,  which we term {\it anomalous}
scaling, is the signature of a departure from the {\it iid} case.
 
We have examined the  empirical behavior of $\sigma(N\tau)$ and $\kappa(N\tau)$: 
 Figure 2. illustrates the scaling behavior of the variance; it can clearly be 
seen that the variance is linear in the timescale i.e.  $\sigma(N\tau) = N\sigma(\tau)$, 
as it would be for independent increments. This property illustrates the
the  additivity of the variances of increments, which is equivalent to the
absence of autocorrelation, discussed below. 

However if we examine the scaling behavior for higher cumulants of the distribution
$\P$ we notice that their behavior is different than for an {\it iid} random walk.
Figure 3 shows the behavior of the kurtosis under a change of timescale.
As indicated above, in the case of {\it iid} increments, the kurtosis decreases
by a factor $1/N$ when we multiply the time scale by $N$ : 
$\kappa(N\tau) = \kappa(\tau)/N $.
In Figure 3 we have given a power-law fit of the kurtosis $\kappa(N\tau)$
as a function of the time scale $T = N\tau$. We find that $\kappa(N\tau)$ decreases
more slowly than $1/N$: $$\kappa(N\tau) \simeq \kappa(\tau)/N^{\alpha}$$ where 
$\alpha \simeq 0.5$.\\

In principle one could calculate higher order cumulants of the distribution and 
examine their scaling properties. However, the error bar for such calculations
increases drastically as we go higher in the order of the cumulants. 

\section{Correlation and dependence}

It is a well-known fact that price movements in liquid markets
do not exhibit any significant autocorrelation: the autocorrelation function of the price changes

$$ C(T) = \frac{\ov{ \d_{t}\d_{t+T}} -  \ov{ \d_{t}}\  \ov{ \d_{t+T}}}{var (\d_t)} $$
  
rapidly decays to zero in a few minutes: for  $T\geq15$ minutes it can be safely  assumed to
be zero within estimation errors\cite{comment}. The absence of economically significant linear correlations in
price increments and asset returns has been widely documented (see \cite{pagan}
and references within) and often cited
as support for the "Efficient Market Hypothesis" \cite{fama}. The fast decay
of the correlation function implies the additivity of variances: for
uncorrelated variables, the variance of the sum is the sum of the variances.
The absence of linear correlation is thus consistent with the linear increase
of the variance with respect to time scale.

However, the absence of serial correlation does not imply the independence of the increments:
it is well known that two random variables may have zero correlation yet
not be independent. In the same way, one may give examples of
 stochastic processes with uncorrelated but not independent increments. One
example, given in \cite{doob} is the following:  let $X$ be a random variable
uniformly distributed on $[0,\pi]$ and  define the discrete time stochastic process
$(U_t)_{t\geq 0}$ by the recurrence relation $U_t = U_{t-1} + sin(tX)$. A simple 
calculation then shows that the increments of $(U_t)_{t\geq 0}$ are uncorrelated;
however, they are not independent: indeed, they are deterministic functions
of the same random variable, $X$.

The  simplest non-parametric test of the {\it iid} hypothesis
is the sign test: if the increments have the same median value
$m$ then the number T of values exceeding $m$ in a sample of size $n$
has expected value $n/2$ and the statistic $s =\frac{2T-n}{\sqrt{n}}$
has a standard normal distribution for large samples. The median $m$
is taken to be the sample median, zero in this case.

We calculate the sign statistic
$$
s =\frac{2T-n}{\sqrt{n}}
$$

for tick data, 5 minute price changes and 30 min price changes. The interval between ticks is irregular and depends on market activity but it is less than
one minute. The results are the following:
$$
s_{tick} = 6.66\ ,\ s_{5min} = -2.1\ , \ s_{30min} = 1.25\  
$$
Assuming that the sign statistic has a N(0,1) distribution and using  a  confidence interval of two standard deviations, the results show that although  the signs 
of price movements at very short time resolutions (less than a minute) are 
significantly correlated, they can be safely considered as independent
(as far as the sign test is concerned) for time scales  beyond 30 minutes.  
Even for 5 minutes $s$ is close to 2 standard deviations.

From a formal point of view, 
independence is  characterized by the fact that for any (measurable)
functions $f$,$g$ the quantity 

$$ C_{f,g}(X,Y) = E [f(X)g(Y)] - E[f(X)] E[g(Y)] $$

vanishes i.e. the variables $f(X)$ and $g(Y)$ are uncorrelated for any non-linear
functions $f$ and $g$. In particular, $C_{f,f}(X,Y)=0$ : applying any non-linear
function to an independent sequence gives an uncorrelated sequence.
In our case this means that increments are stationary 
and independent if and only if for any function $f$ the series $( f\dx )_{t\geq0}$ 
is serially uncorrelated. This criterion indicates a simple non-parametric
way of testing the hypothesis of {\it iid} increments: for a given function
$f$ we calculate  

$$ C_{f}(T) = \frac{\ov{ f(\d_t) f(\d_{t+T})} - \ov{f(\d_{t})}\ \ov{f(\d_{t+T})}}{var\ f(\d_t)} $$ 

Formally, $C_f(T)$ corresponds to a resummation of
a certain subclass of correlation functions of the increments $(\d_t)_{t\geq0}$, with weights
corresponding to the Taylor coefficients of $f$. We use the following non-linear
test functions: $f_1(x) = x^2 , f_2(x) = cos(x)$ and $f_3(x) = ln(1+x^2)$.
In each case we calculate the autocorrelation of the series $( f (\d_t))_{t\geq0}$.
 The results are displayed in Figure 5.
It can be seen that the series $(f_1(\d_t))_{t\geq0}$ and $(f_3(\d_t))_{t\geq0}$ exhibit 
slowly decaying serial correlations, showing the presence of nonlinear
dependence in the data.

To obtain a more detailed picture
of the dependence structure of the increments, we proceed to calculate
higher order correlation functions. It is well known that the square of
the returns  -or any other measure of the amplitude of fluctuations for
example the absolute value of the increments-  
exhibits significant autocorrelation. 

Figure 6 compares the autocorrelation of the price changes to that of their absolute value. In contrast to the autocorrelation function of the increments
which decays rapidly to zero in a few minutes, the autocorrelation of their 
absolute value decays slowly to zero while staying  positive, indicating 
persistence in the scale of the fluctuations, a phenomenon which can be related
to the well known "clustering of volatility".

Another measure of the scale of the fluctuations os given by the square
of the increments; Figure 7 displays the autocorrelation function $g(T)$
of the square of the increments, defined as:

$$
g(N) = g(\frac{T}{\tau}) = \frac{\ov{\d_t^2 \d_{t+T}^2 } - \ov{\d_t^2}\ \ov{ \d_{t+T}^2 }}{var (\d ^2)} = \frac{\ov{\d_t^2 \d_{t+T}^2 } - \ov{\d_t^2}\ \ov{ \d_{t+T}^2 }}
{\mu_4(\tau) - \sigma(\tau)^4} \eqno{(1)}$$

Fitting $g(T)$ by an exponential gives very bad results; however the slow decay $g(T)$ is well  represented by a power law:

$$
g(k) \simeq \frac{g_0}{k^\alpha}\ \ \alpha = 0.37 \pm 0.037 \ g_0 = 0.08
$$

the exponent $\alpha$ and the constant $g_0$ being obtained
by a regression of $\ln{g(T)}$ against $T$.   We will see in the following section that this approximation is precise enough to finely capture
the scaling behavior of the kurtosis.

\section{Relation between scaling behavior and correlation structure}
 
The results of the two preceding sections may be blended into a consistent picture
of the dynamical properties of the price process by  remarking that the scaling behavior
of the semi-invariants of the increments is related to their correlation structure.

There is a simple, well-known example of relations between correlation functions and scaling behavior: the relation between the variance of a sum
and the covariance of the addends. The variance at time scale $N\tau$ may be expressed
 as 

\begin{eqnarray*}
\sigma^2(N\tau) &=& \sum_{k=1}^{N} \sigma^2(\tau) + 2  \sigma^2(\tau)\sum_{k>l}  C( (k - l) \tau )\\
& = & \sum_{k=1}^{N} \sigma^2(\tau) + 2  \sigma^2(\tau)\sum_{k=1}^{N} (N-k) C( k   \tau )\\ 
&=& N \sigma^2(\tau) [ 1 + 2\sum_{k=1}^{N} (1-\frac{k}{N}) C( k   \tau ) ]\\
\end{eqnarray*}

where $C(T)$ is the autocorrelation function of the increments, defined above.
  therefore implies $\sigma^2(N\tau) = N\sigma^2(\tau) $. The absence of autocorrelation
implies linear scaling of the variance.

A similar relation may be derived between the scaling behavior of the kurtosis
and the autocorrelation of the square of the increments (the function $g$
defined above)\cite{adapt}.   
Using the results above, we model the signs of the successive price
changes as an independent  sequence with zero mean, independent from their
absolute value. Using these hypotheses the fourth moment  may be expressed in terms of the moments and
correlation functions of shorter time scale increments $\d$; the
calculation, details of which are given in the appendix, gives:

\begin{eqnarray*}
\mu_4(N\tau) & =  & N\mu_4 (\tau) + 3  (N^2 + N)\sigma(\tau)^4 + \\
&  & 6 N (\mu_4 - \sigma(\tau)^4) \sum_{k=1}^{N}  g(k) - 6(\mu_4 - \sigma(\tau)^4) \sum_{k=1}^{N} k g(k)\\
\end{eqnarray*}

The scaling behavior of the kurtosis is therefore  related
to the behavior of the correlation function $g$; if $g$ exhibits a slow
power law decay $g(k) \simeq \frac{g_0}{k^\alpha}$ then $\sum_{k=1}^{N}  g(k) \simeq g_0/(1-\alpha)N^{1-\alpha}$ and $\sum_{k=1}^{N} k g(k) \simeq g_0/(2-\alpha)N^{2-\alpha}$ 
so:

\begin{eqnarray*}
\kappa(N\tau) & =  & \frac{\mu_4 (N\tau)}{\sigma^4(N\tau)} - 3 \\
			& = & \frac{\kappa(\tau)}{N} + 
\frac{ 6(\kappa(\tau)+2) }{ (2-\alpha) (1-\alpha) N^{\alpha}}
\end{eqnarray*}  

Figure 8. shows good agreement for large values of N ($N\geq 50$) between the right hand side of the above equation  and the kurtosis as a function of time resolution, plotted on the same diagram. We have thus shown that the anomalous 
scaling of the kurtosis is well accounted for by the presence of correlations
in the scale of the fluctuations, represented by the square of the increments.

\section{Conclusion}

By comparing the statistical properties of price increments
of S\&P index futures at various time scales, we have shown that
 studying the resolution dependence or {\it scaling behavior}
of the variance and the kurtosis of the empirical distribution
of price increments  enables us to extract much more information
than studies on a given time scale. In particular, the scaling
properties of the kurtosis of the price changes may be recovered from the correlation
function of their squares, which exhibits a slow power law decay with exponent 
$\alpha = 0.37 \pm 0.037$. Such relations may be generalized 
to other cumulants and moments but which are less interesting
from an economic point of view.\\

\noindent
$\bullet$ Acknowledgements:\\

The author is grateful to Jean-Philippe Bouchaud and Marc Potters for their 
constant help and helpful comments and {\it Science \& Finance } for providing the data.
 
\vskip 1cm

{\bf Appendix : relation between scaling behavior of kurtosis and correlation of increment squares }
\vskip 1cm

 Let $\d_{k} = x(t +k \tau) - x(t+(k-1)\tau) $. We derive here the relation between the
kurtosis   and the correlation function of the square of the increments
$g(k)$ defined in equation (1). Let us furthermore suppose without
loss of generality that $E[\d_{k}=0]$ (see also the remark in section 3).
The fourth moment
$\mu_4 (N\tau)$ is given by

\begin{eqnarray*}
\mu_4 (N\tau) & = & E [ ( x(t +N\tau) - x(t) )^4 ] \\
 & = & E [ (\sum_{k=1}^{N} \d_{k}) ^4 ]  \\
       & = &  E [ \sum_{i,j,k,l=1}^{N} \d_{i} \d_{j} \d_{k} \d_{l} ]\\
      & = &  \sum_{i=1}^{N}E [  \d_{i} ^4 ] + 6 \sum_{i>j}  E [ \d_{i}^2 \d_{j}^2 ] + 6 \sum_{i>j >k=1}^{N}   E [ \d_{i}^2 \d_{j}\d_{k} ]  \\
      &  & +  3 \sum_{i>k}   E  [ \d_{i}^3 \d_{k} ] + \sum_{i>j>k>l} E [  \d_{i} \d_{j} \d_{k} \d_{l} ]\\
\end{eqnarray*}

We will suppose now that  the increments may be written as
$$ \d_k = \epsilon_k \gamma_k\  \ \epsilon_k \ \in \{ +1,-1 \} ,\  \gamma_k >0$$
where $(\epsilon_k)$ is a sequence of independent sign variables, but the sequence
 $(\gamma_k)$ may have an arbitrary dependence structure. Since $\d_k$ is centered
$E \epsilon_k = 0$. This hypothesis  captures well
the behavior of the increments: their magnitudes are correlated ("ARCH effect")
 but their signs random. 
We furthermore suppose that $\gamma_k$ is independent from $\epsilon_j$. As a result, most of the terms appearing in the above sum vanish:

\begin{eqnarray*}
 E  [ \d_{i}^3 \d_{k} ]  & = & E  [ \epsilon_i  \epsilon_k \gamma_{i}^3 \gamma_{k} ] \\
& = &  E  [ \epsilon_i]  E [ \epsilon_k] . E  [ \gamma_i^3 \gamma_k] = 0\\
E [  \d_{i} \d_{j} \d_{k} \d_{l} ] & = & E  [ \epsilon_i] E[ \epsilon_j] E[ \epsilon_k]   E[ \epsilon_l] E  [ \gamma_i \gamma_j \gamma_k \gamma_l]=0\\
E [  \d_{i}^2 \d_{j} \d_{k}  ] & = &  E[ \epsilon_j] E[ \epsilon_k] E [ \gamma_i^2 \gamma_j \gamma_k] = 0\\
\end{eqnarray*}

 The expression for the fourth moment therefore reduces to:

\begin{eqnarray*}
\mu_4 (N\tau) & = & \sum_{i=1}^{N}E [  \d_{i\tau} ^4 ] + 6 \sum_{i>j}  E [ \d_{i\tau}^2 \d_{j\tau}^2 ]\\
& = & \sum_{j=1}^{N}E [  \d_{j\tau} ^4 ] + 6 \sum_{j=1}^{N} (N-j)  
E [ \d_{t}^2 \d_{t+j\tau}^2 ]\\
\end{eqnarray*}

From (1), $E [\d_{i}^2 \d_{j}^2]$ may now be expressed in terms of the correlation function $g$:

$$
E [ \d_{i\tau}^2 \d_{j\tau}^2 ] = (\mu_4 - \sigma(\tau)^4) \ g(|i-j|) + \sigma(\tau)^4
$$
hence, by substitution, the relation between $\mu_4$ and $g$:

\begin{eqnarray*}
\mu_4 (N\tau)& = & N\mu_4 (\tau) +  6 \sum_{k=1}^{N} (N-k) [ \sigma(\tau)^4 + 
(\mu_4 - \sigma(\tau)^4) g(k) ]\\
& = & N\mu_4 (\tau) + 3  (N^2 - N)\sigma(\tau)^4 + \\
& = & 6 N (\mu_4(\tau) - \sigma(\tau)^4) \sum_{k=1}^{N}  g(k) - 6(\mu_4(\tau) - \sigma(\tau)^4) \sum_{k=1}^{N} k g(k)\\
\end{eqnarray*}

The kurtosis for a time resolution $T= N\tau$ is therefore given by: 

\begin{eqnarray*}
\kappa(N\tau) & =  & \frac{\mu_4 (N\tau)}{\sigma(N\tau)^4} - 3 \\
			& = & \frac{N\mu_4 (\tau) }{N^2\sigma(\tau)^4} 
+ 3(1-\frac{1}{N}) - 3 +  6(\mu_4(\tau) - \sigma(\tau)^4)\sum_{k=1}^{N}\frac{(N-k) g(k)}{N^2\sigma^4(\tau)} \\
& = &\frac{\kappa(\tau)}{N} + \frac{6(\kappa(\tau)+2)}{N}\sum_{k=1}^{N} (1-\frac{k}{N}) g(k) \\
\end{eqnarray*}   

In our case $g(T) \simeq \frac{g_0}{T^\alpha} $ which implies, for $\kappa(T)$:

\begin{eqnarray*}
\kappa(N\tau) & \simeq  & \frac{\kappa(\tau)}{N} + \frac{6 g_0 (\kappa(\tau)+2)}{N^2}(\frac{N^{2-\alpha}}{1-\alpha}-\frac{N^{2-\alpha}}{2-\alpha})\\
  & = & \frac{\kappa(\tau)}{N} + 
\frac{ 6 g_0 (\kappa(\tau)+2) }{ (2-\alpha) (1-\alpha) N^{\alpha} }\\
\end{eqnarray*}   

\vskip 1cm
\bibliographystyle{unsrt}

\newpage

FIGURE CAPTIONS:
\vskip 1cm
\noindent
Figure 1: Probability density of 5 minute increments of S\&P 500 index future prices. The lower curve is a gaussian with same mean and variance.
\vskip 1cm
\noindent
Figure 2: Scaling behavior of the variance of price increments: the variance
increases approximately linearly with the time scale, which is consistent with the absence
of significant linear correlations.
\vskip 1cm
\noindent
Figure 3: Scaling behavior of the kurtosis of price increments. The kurtosis decreases much more slowly than 1/N: a power law fit gives an exponent close to 0.5 .
\vskip 1cm
\noindent
Figure 4: Autocorrelation function of price changes. Time lag is given
in multiples of $\tau = $ 5 minutes.
\vskip 1cm
\noindent
Figure 5: Behavior of some non-linear correlation functions of price changes.
\vskip 1cm
\noindent
Figure 6: Autocorrelation function of absolute price changes: contrary to 
the price changes, their absolute values display a persistent character. 
\vskip 1cm
\noindent
Figure 7: Autocorrelation function of the square of price changes.
Note the slow decay, well represented by a power law with exponent $\alpha=0.37$. 
\vskip 1cm
\noindent
Figure 8: The scale dependence of the kurtosis (data points) may be reconstructed (solid line) from the autocorrelation function of the square of the price increments, approximated by a power law (Eq.1).
\vskip 1cm
\noindent
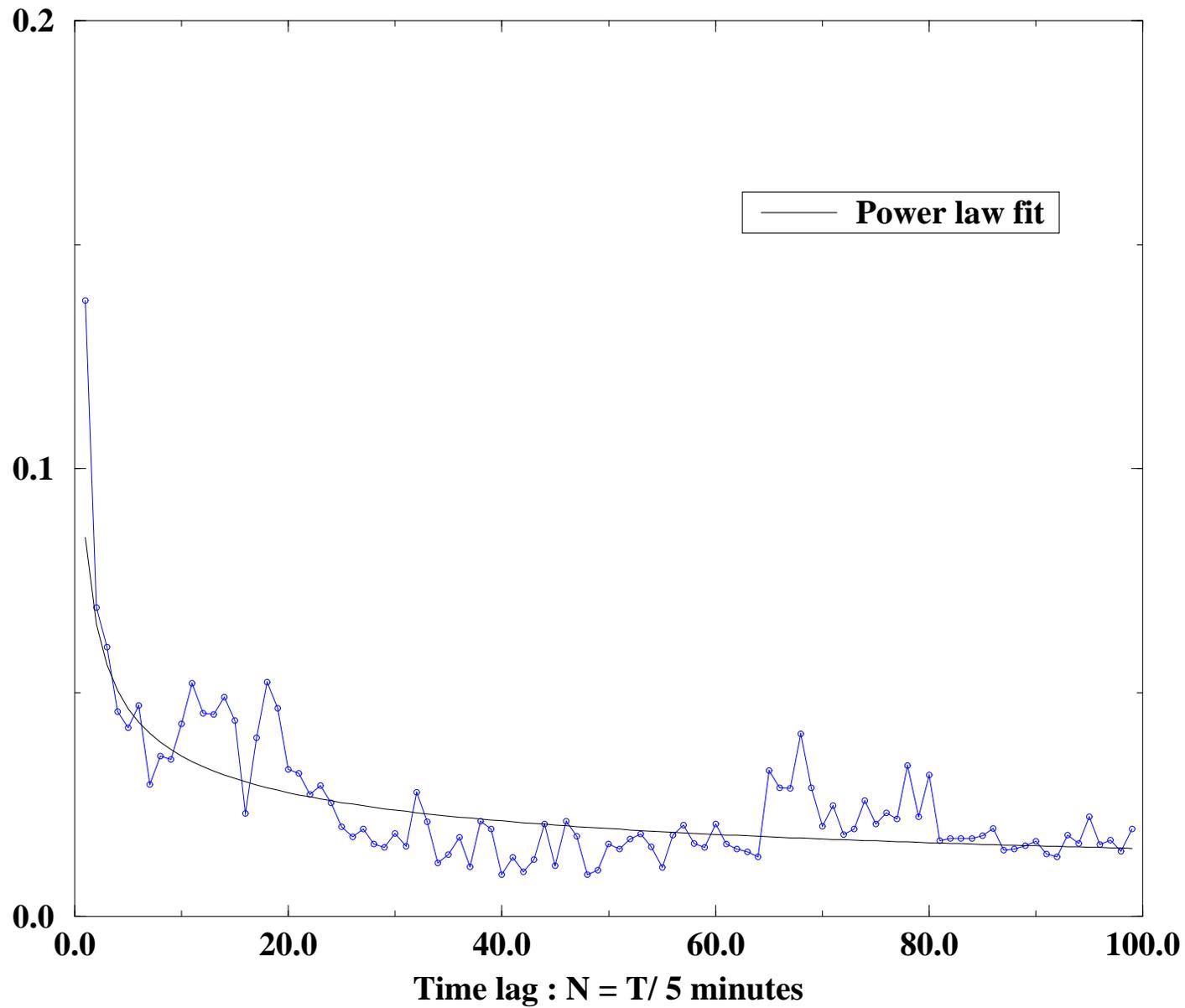
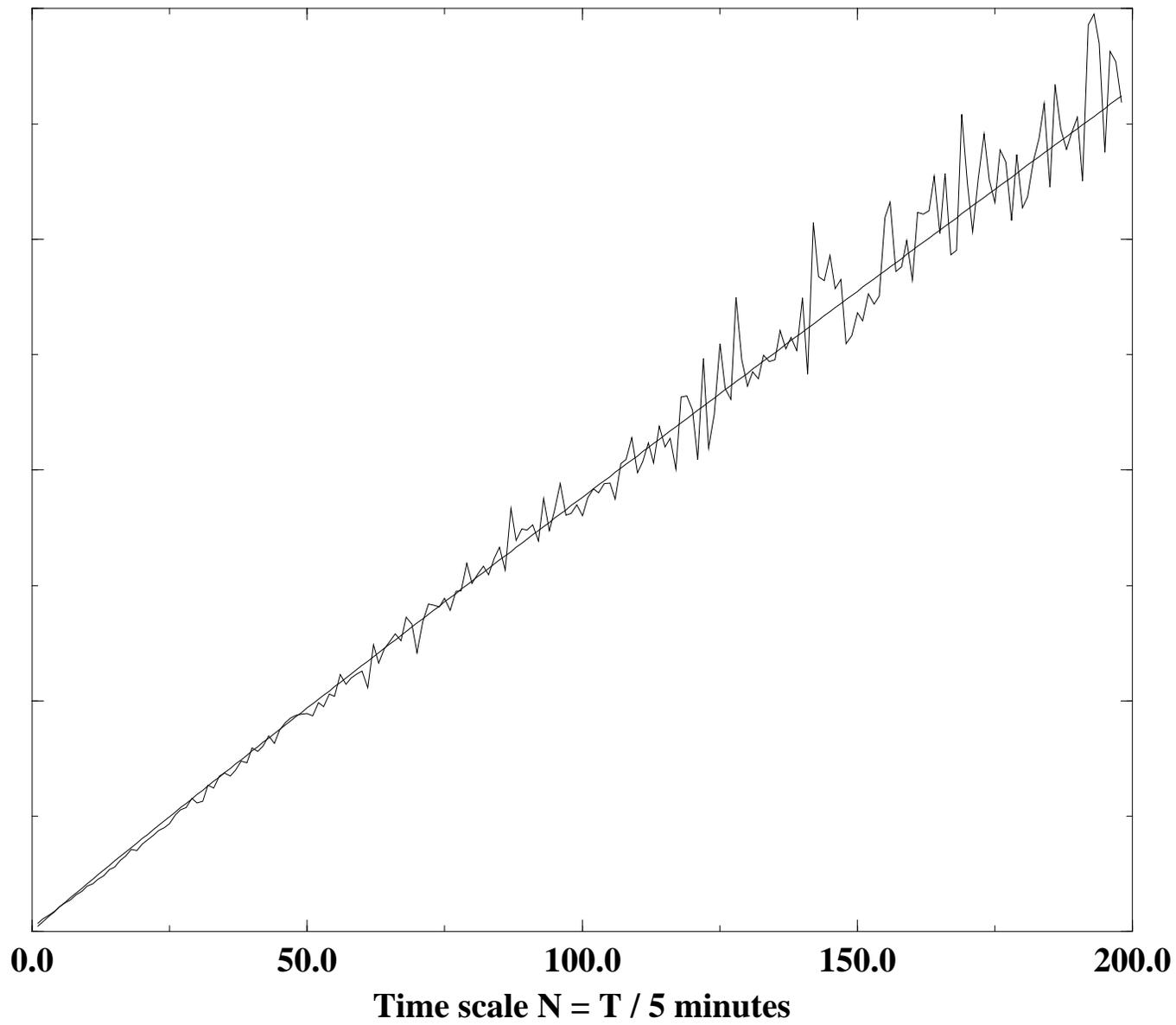
\end{document}